\documentclass[
    ,final            
  ]
  {aipproc}

\layoutstyle{6x9}

\begin{document}

\title{Cosmic star formation history \\revealed by the AKARI \\  {\small \& Spatially-resolved spectroscopy of an E+A (Post-starburst) system}}

\classification{98.70.Lt}
\keywords      {galaxies: evolution, galaxies:starburst}

\author{Tomotsugu GOTO}{
  address={Institute for Astronomy, University of Hawaii, 2680 Woodlawn Drive, Honolulu, HI, 96822, USA}
}

\author{the AKARI NEPD team}{
  address={Japan Aerospace Exploration Agency, Sagamihara, Kanagawa 229-8510, Japan}
}

\author{M.Yagi}{
  address={National Astronomical Observatory, 2-21-1 Osawa, Mitaka, Tokyo, 181-8588,Japan}
}

\author{C.Yamauchi}{
  address={Japan Aerospace Exploration Agency, Sagamihara, Kanagawa 229-8510, Japan}
}

\begin{abstract}
 We reveal cosmic star-formation history obscured by dust using deep infrared observation with the AKARI. A continuous filter coverage in the mid-IR wavelength (2.4, 3.2, 4.1, 7, 9, 11, 15, 18, and 24$\mu$m) by the AKARI satellite allows us to estimate restframe 8$\mu$m and 12$\mu$m luminosities without using a large extrapolation based on a SED fit, which was the largest uncertainty in previous work. 
We found that restframe 8$\mu$m ($0.38<z<2.2$), 12$\mu$m ($0.15<z<1.16$), and  total infrared (TIR) luminosity functions (LFs) ($0.2<z<1.6$) constructed from the AKARI NEP deep data, show a continuous and strong evolution toward higher redshift. 
 In terms of cosmic infrared luminosity density ($\Omega_{IR}$), which was obtained by integrating analytic fits to the LFs, we found a good agreement with previous work at $z<1.2$, with $\Omega_{IR}\propto (1+z)^{4.4\pm 1.0}$.
 When we separate contributions to $\Omega_{IR}$ by LIRGs and ULIRGs, we found more IR luminous sources are increasingly more important at higher redshift.  We found that the ULIRG (LIRG) contribution increases by a factor of 10 (1.8) from $z$=0.35 to $z$=1.4.
\end{abstract}

\maketitle


\paragraph{{\bfseries{Introduction}}}
Revealing the cosmic star formation history is one of the major goals of the observational astronomy. However, UV/optical estimation only provides us with a lower limit of the star formation rate (SFR) due to the obscuration by dust. 
A straightforward way to overcome this problem is to observe in infrared, which can capture the star formation activity invisible in the UV. 
The superb sensitivities of recently launched Spitzer and AKARI satellites can revolutionize the field.

However, most of the Spitzer work relied on a large extrapolation from 24$\mu$m flux to estimate the 8, 12$\mu$m or total infrared (TIR) luminosity, due to the limited number of mid-IR filters.
AKARI has continuous filter coverage across the mid-IR wavelengths,  thus, allows us to estimate mid-IR luminosity without using a large $k$-correction based on the SED models, eliminating the largest uncertainty in previous work. 
By taking advantage of this, we present the restframe 8, 12$\mu$m TIR LFs, and thereby the cosmic star formation history derived from these using the AKARI NEP-Deep data.
 

\begin{figure}
\includegraphics[scale=0.35]{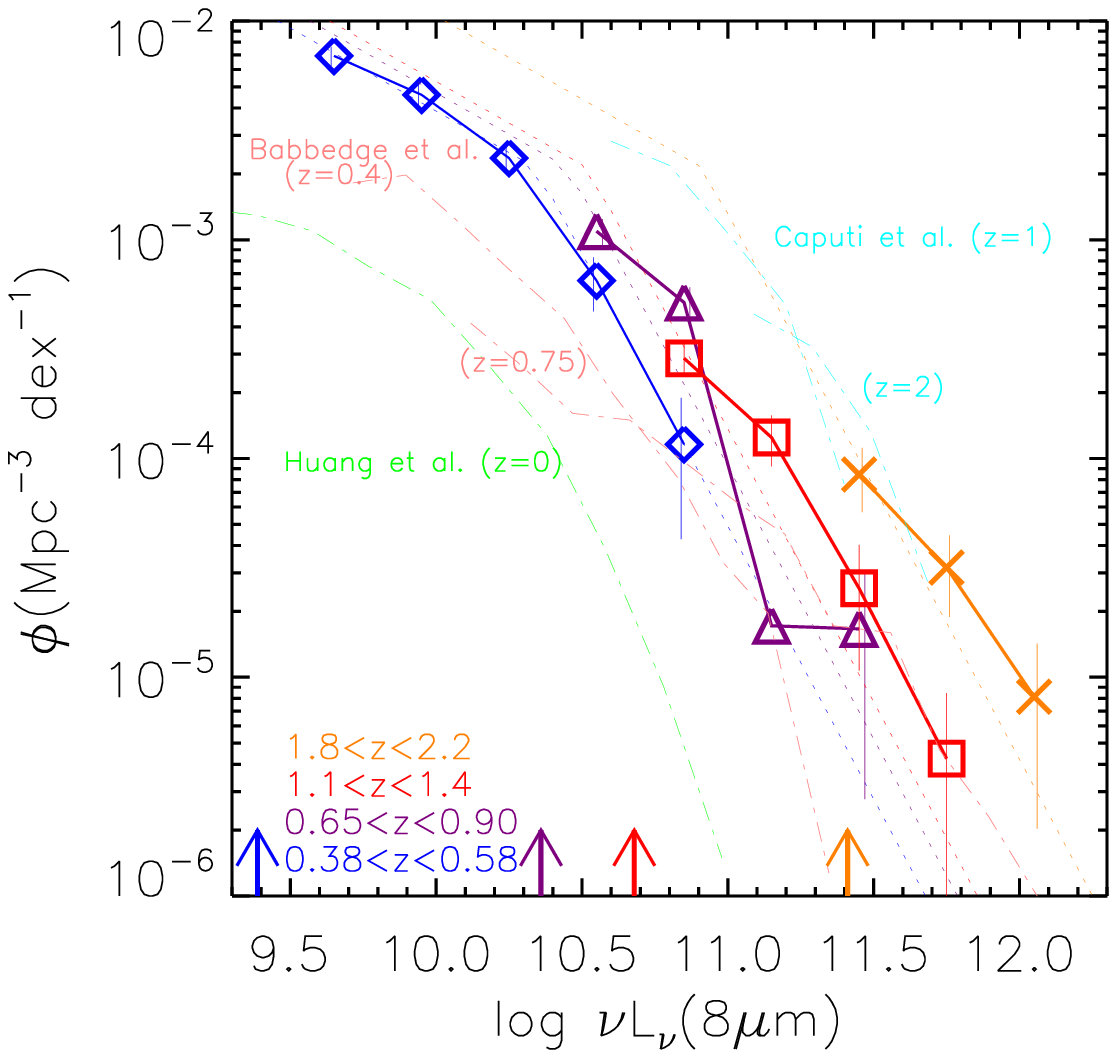}
\includegraphics[scale=0.35]{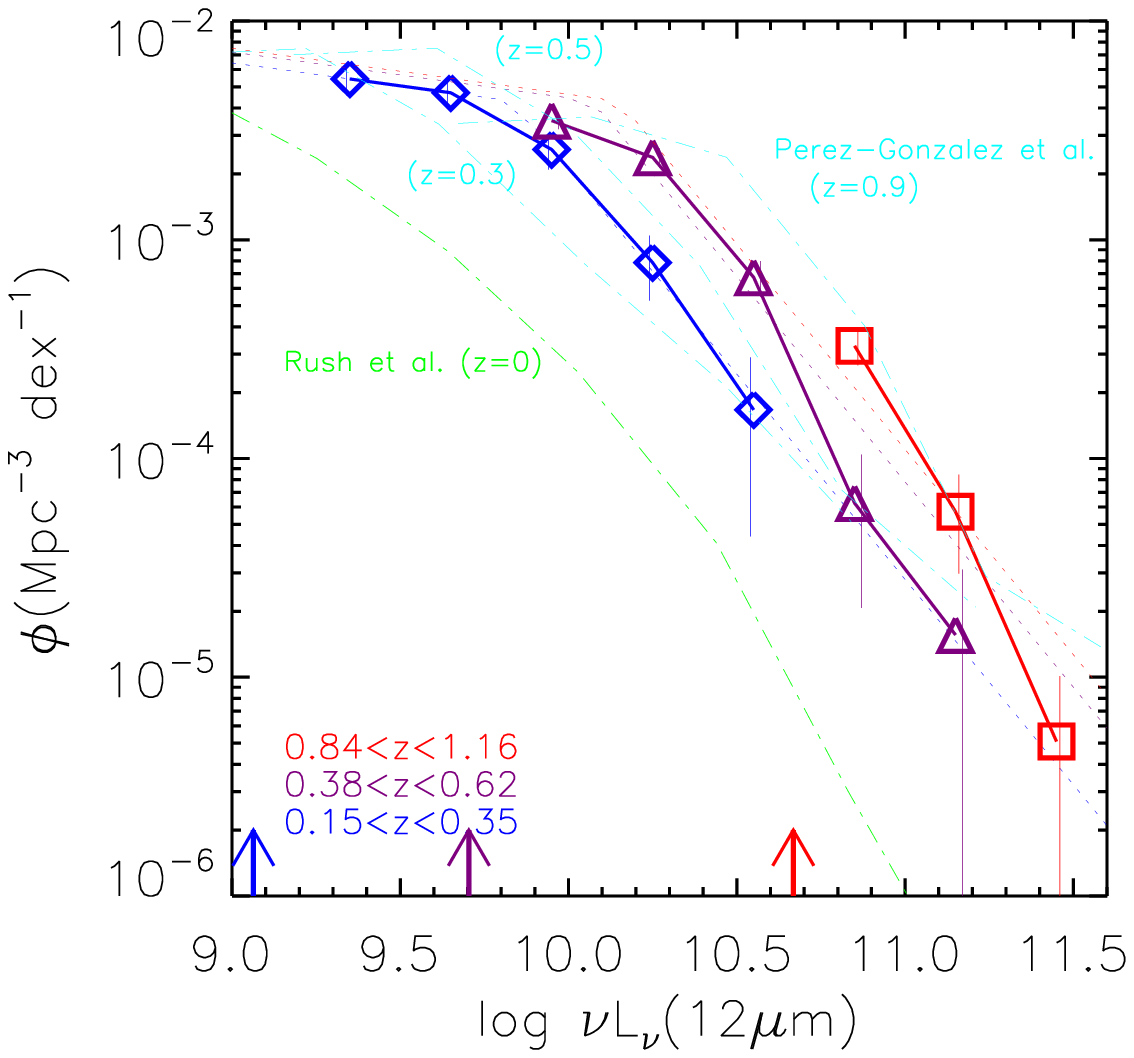}
\includegraphics[scale=0.35]{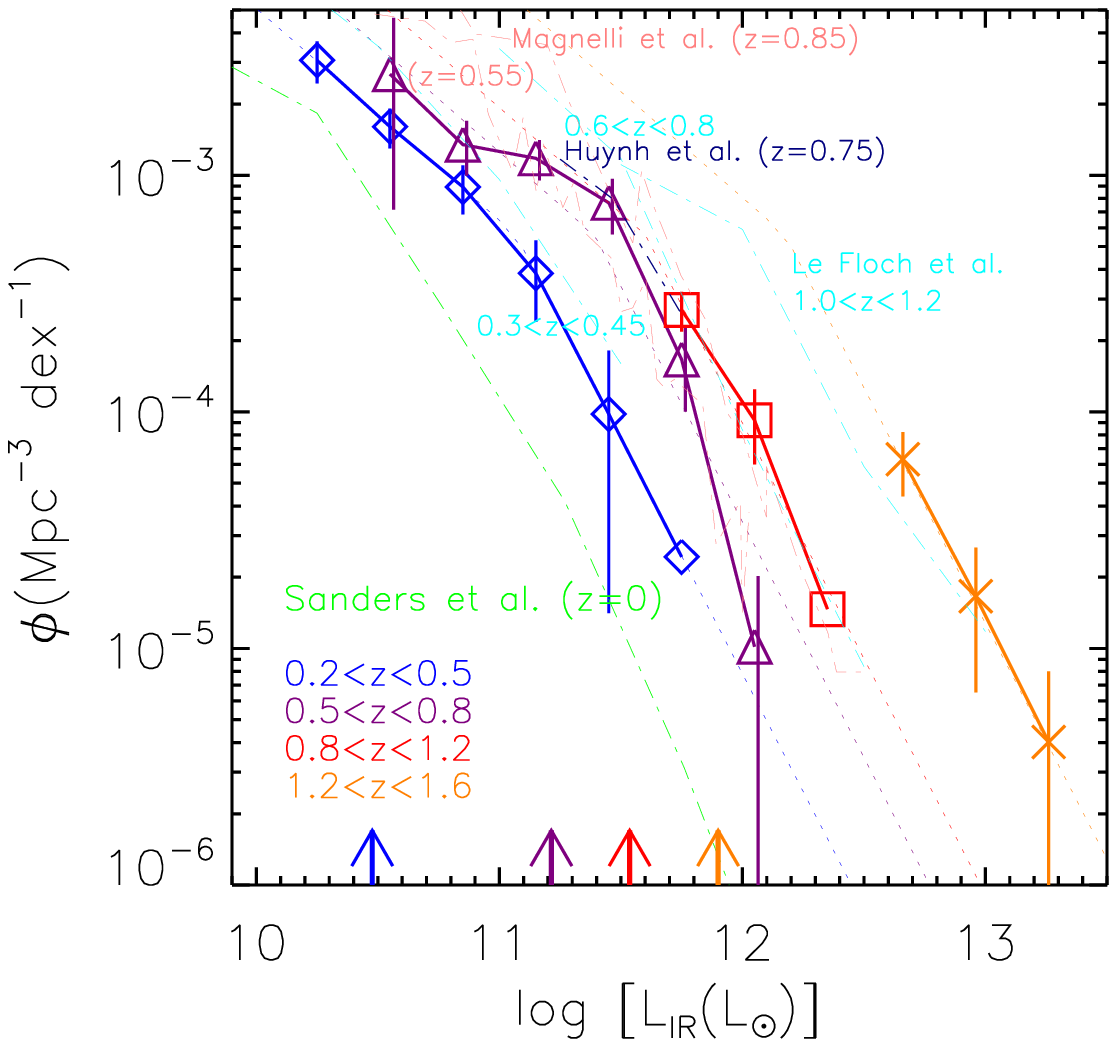}
\caption{
(left) Restframe  8$\mu$m LFs.
 The blue diamonds, purple triangles, red squares, and orange crosses show the 8$\mu$m LFs at $0.38<z<0.58, 0.65<z<0.90, 1.1<z<1.4$, and $1.8<z<2.2$, respectively. 
 The dotted lines show analytical fits with a double-power law.
 Vertical arrows show the 8$\mu$m luminosity corresponding to the flux limit at the central redshift in each redshift bin.
 Overplotted are  \citet{2006MNRAS.370.1159B} in the pink dash-dotted lines, \citet{2007ApJ...660...97C} in the cyan dash-dotted lines, and \citet{2007ApJ...664..840H} in the green dash-dotted lines. AGNs are excluded from the sample.
(middle)
Restframe  12$\mu$m LFs.
  The blue diamonds, purple triangles, and red squares show the 12$\mu$m LFs at $0.15<z<0.35, 0.38<z<0.62$, and $0.84<z<1.16$, respectively.
  Overplotted are  \citet{2005ApJ...630...82P} at $z$=0.3,0.5 and 0.9 in the cyan dash-dotted lines, and \citet{1993ApJS...89....1R} at $z$=0 in the green dash-dotted lines.
(right)
TIR LFs.
}\label{fig:8umlf}
\end{figure}

\paragraph{{\bfseries Data \& Analysis}}

The AKARI has observed the NEP deep field (0.4 deg$^2$) in 9 filters ($N2,N3,N4,S7,S9W,S11,L15,L18W$ and $L24$) to the depths of 14.2, 11.0, 8.0, 48, 58, 71, 117, 121 and 275$\mu$Jy (5$\sigma$)\citep{2008PASJ...60S.517W}. 
This region is also observed in $BVRi'z'$ (Subaru),  $u'$ (CFHT), $FUV,NUV$ (GALEX), and $J,Ks$ (KPNO2m), with which we computed photo-z with $\frac{\Delta z}{1+z}$=0.043.  Objects which are better fit with a QSO template are removed from the analysis.
We compute LFs using the 1/$V_{\max}$ method. Data are used to 5$\sigma$ with completeness correction. Errors of the LFs are from 1000 realization of Monte Carlo simulation.


\begin{figure}
  \includegraphics[height=.4\textheight]{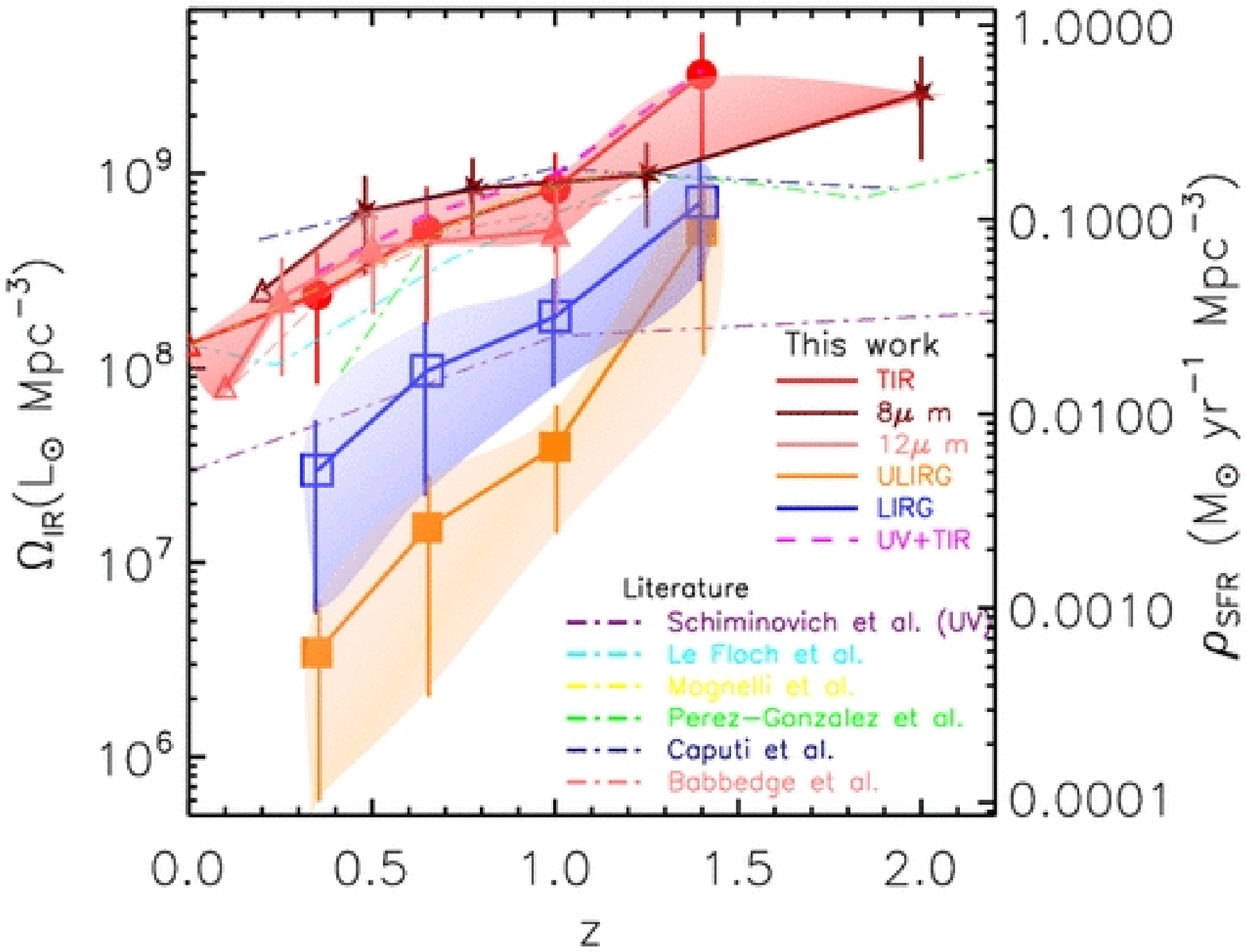}
  \caption{Evolution of TIR luminosity density based on TIR LFs (red circles), 8$\mu$m LFs (stars), and 12$\mu$m LFs (filled triangles). The blue open squares and orange filled squares  are for LIRG and ULIRGs only, also based on our $L_{TIR}$ LFs.
Overplotted dot-dashed lines are estimates from the literature: \citet{2005ApJ...632..169L}, \citet{2009A&A...496...57M} , \citet{2005ApJ...630...82P}, \citet{2007ApJ...660...97C},   and \citet{2006MNRAS.370.1159B} are in cyan, yellow, green, navy, and pink, respectively.
The purple dash-dotted line shows UV estimate by \citet{2005ApJ...619L..47S}.
The pink dashed line shows the total estimate of IR (TIR LF) and UV \citep{2005ApJ...619L..47S}.}\label{fig:TLD_all}
\end{figure}
\paragraph{{\bfseries 8$\mu$m LF}}

Monochromatic 8$\mu$m luminosity ($L_{8\mu m}$) is known to correlate well with the TIR luminosity \citep{2006MNRAS.370.1159B,2007ApJ...664..840H}, especially for star-forming galaxies because the rest-frame 8$\mu$m flux are dominated by prominent PAH features such as at 6.2, 7.7 and 8.6 $\mu$m.
 The left panel of Fig.\ref{fig:8umlf} shows a strong evoltuion of 8$\mu$m LFs.
 Overplotted previous work had to rely on SED models to estimate $L_{8\mu m}$ from the Spitzer $S_{24\mu m}$ in the MIR wavelengths where SED modeling is difficult due to the complicated PAH emissions. Here, AKARI's mid-IR bands are advantageous in  directly observing redshifted restframe 8$\mu$m flux in one of the AKARI's filters, leading to more reliable measurement of 8$\mu$m LFs without uncertainty from the SED modeling. 

\paragraph{{\bfseries 12$\mu$m LF}}

 12$\mu$m luminosity ($L_{12\mu m}$) represents mid-IR continuum, and known to correlate closely with TIR luminosity \citep{2005ApJ...630...82P}. 
  The middle panel of Fig.\ref{fig:8umlf} shows a strong evoltuion of 12$\mu$m LFs.
 Here the agreement with previous work is better because (i)  12$\mu$m continuum is easier to be modeled, and (ii) the Spitzer also captures restframe 12$\mu$m in $S_{24\mu m}$ at z=1.

\paragraph{{\bfseries TIR LF}}
 Lastly, we show the TIR LFs in the right panel of Fig.\ref{fig:8umlf}. 
We used \citet{2003MNRAS.338..555L}'s SED templates to fit the photometry using the AKARI bands at $>$6$\mu$m ($S7,S9W,S11,L15,L18W$ and $L24$). 
 The TIR LFs show a strong evolution compared to local LFs. 
 At $0.25<z<1.3$, $L^*_{TIR}$ evolves as $\propto (1+z)^{4.1\pm0.4}$.


\paragraph{{\bfseries Cosmic star formation history}}

We fit LFs in Fig.\ref{fig:8umlf} with a double-power law, then integrate to estimate total infrared luminosity density at various z. The restframe 8 and 12$\mu$m LFs are converted to $L_{TIR}$ using \cite{2005ApJ...630...82P,2007ApJ...660...97C} before integration.
 The resulting evolution of the TIR density is shown in Fig.\ref{fig:TLD_all}.
 The right axis shows the star formation density assuming \citet{1998ARA&A..36..189K}. 
 We obtain $\Omega_{IR}(z) \propto (1+z)^{4.4\pm 1.0}$.
Comparison to $\Omega_{UV}$ \citep{2005ApJ...619L..47S} suggests that $\Omega_{TIR}$ explains 70\% of $\Omega_{total}$ at $z$=0.25, and that by $z$=1.3, 90\% of the cosmic SFD is explained by the infrared. This implies that $\Omega_{TIR}$ provides good approximation of the  $\Omega_{total}$ at $z>1$.
 
 In Fig.\ref{fig:TLD_all}, we also show the contributions to $\Omega_{TIR}$ from LIRGs and ULIRGs.
 From $z$=0.35 to $z$=1.4, $\Omega_{IR}$ by LIRGs increases by a factor of $\sim$1.6, and 
  $\Omega_{IR}$ by ULIRGs increases by a factor of $\sim$10.
More details are in \citet{Goto2010}.


\begin{figure}
 \includegraphics[height=.3\textheight]{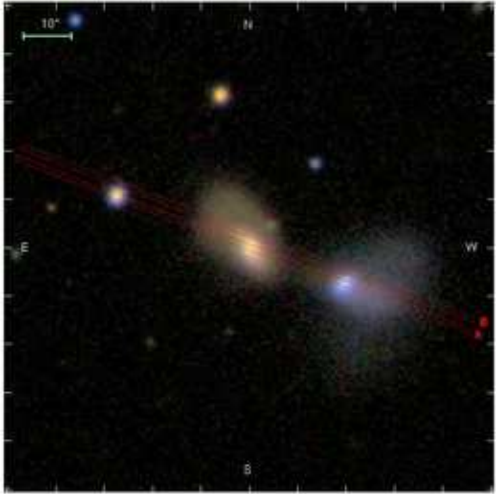}
  \includegraphics[height=.3\textheight]{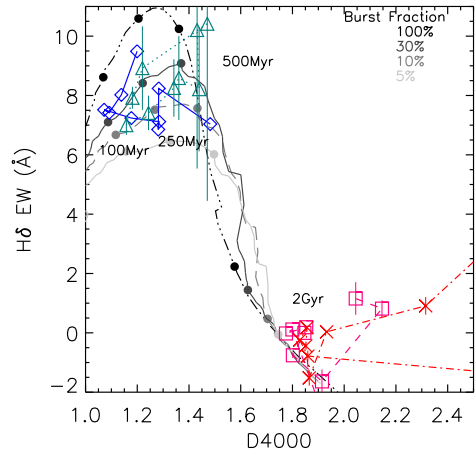}

  \caption{
(left)
The SDSS $g,r,i$-composite image of the J1613+5103. The long-slit positions are overlayed. 
 The E+A galaxy is to the right (west), with bluer colour. The companion galaxy is to the left (east).
(right)
H$\delta$ EW is plotted against D4000. 
The diamonds and triangles are for the E+A core/north spectra, respectively.  The squares and crosses are for the companion galaxy's core/north spectra. 
Gray lines are population synthesis models with 5-100\% delta burst population added to the 10G-year-old exponentially-decaying ($\tau$=1Gyr) underlying stellar population. Salpeter IMF and metallicity of $Z=0.008$ are assumed. On the models, burst ages of 0.1, 0.25, 0.5 and 2 Gyr are marked with the filled circles.
 }\label{fig:d4000}
\end{figure}

\paragraph{{\bfseries Spatially-Resolved Spectroscopy of an E+A (post-starburst) System}}

  We performed a spatially-resolved medium resolution long-slit spectroscopy of 
 a nearby E+A (post-starburst) galaxy system with FOCAS/Subaru \citep{2008MNRAS.391..700G}. 
 This E+A galaxy has an obvious companion galaxy 14kpc in front (Fig.\ref{fig:d4000}, left) with the velocity difference of 61.8 km/s. 

 We found that H$\delta$ equivalent width (EW) of the E+A galaxy is greater than 7\AA~ galaxy wide (8.5 kpc) with no significant spatial variation. 
 We detected a rotational velocity in the companion galaxy of $>$175km/s.  The progenitor of the companion may have been a rotationally-supported, but yet passive S0 galaxy. 
The age of the E+A galaxy after quenching the star formation is estimated to be 100-500Myr, with its centre having slightly younger stellar population.
 The companion galaxy is estimated to have older stellar population of $>$2 Gyrs of age with no significant spatial variation (Fig.\ref{fig:d4000}, right).
 
 These findings are inconsistent with a simple picture where the dynamical interaction creates infall of the gas reservoir that causes the central starburst/post-starburst. Instead, our results present an important example where the galaxy-galaxy interaction can trigger a galaxy-wide post-starburst phenomena.

\end{document}